# Origin of isolated olivine grains in carbonaceous chondrites

Emmanuel Jacquet[1], Maxime Piralla[2], Pauline Kersaho[2], Yves Marrocchi[2]

[1]Institut de Minéralogie, de Physique des Matériaux et de Cosmochimie (IMPMC), Muséum national d'Histoire naturelle, Sorbonne Université, CNRS ; CP52, 57 rue Cuvier, 75005 Paris, France.

[2]Centre de Recherches Pétrographiques et Géochimiques, CNRS, Université de Lorraine, UMR 7358, 54501 Vandœuvre-lès-Nancy, France.

E-mail: emmanuel.jacquet@mnhn.fr

*Abstract*

We report microscopic, cathodoluminescence, chemical and O isotopic measurements of FeO-poor isolated olivine grains (IOG) in the carbonaceous chondrites Allende (CV3), Northwest Africa 5958 (C2-ung), Northwest Africa 11086 (CM2-an), Allan Hills 77307 (CO3.0). The general petrographic, chemical and isotopic similarity with *bona fide* type I chondrules confirms that the IOG derived from them. The concentric CL zoning, reflecting a decrease in refractory elements toward the margins, and frequent rimming by enstatite are taken as evidence of interaction of the IOG with the gas as stand-alone objects. This indicates that they were splashed out of chondrules when these were still partially molten. CaO-rich refractory forsterites, which are restricted to $\Delta^{17}O < -4$ ‰ likely escaped equilibration at lower temperatures because of their large size and possibly quicker quenching. The IOG thus bear witness to frequent collisions in the chondrule-forming regions.

# 1. Introduction

In their seminal description of Murchison, Fuchs et al. (1973) reported isolated olivine grains (henceforth IOG), one of which earned the honor of their frontispiece. With the condensation models in full swing since the fall of Allende (e.g., Grossman 1972; Marvin et al. 1970), it was but natural to interpret these grains, soon found in other carbonaceous chondrite clans, as nebular condensation products, for which Olsen and Grossman (1974, 1978) adduced morphological evidence such as large size, euhedral shapes and patterned crystal surfaces. Yet chemical analogies of the IOG with chondrule olivine, as well as the presence of glass inclusions chemically comparable to chondrule mesostases, suggested early that they were chondrule fragments, possibly liberated after alteration of the mesostasis (e.g. McSween (; Richardson and McSween 1978; Desnoyers (; Nagahara and Kushiro (; Jones (1992.

Still, Steele (1986) drew attention to a subset of IOG, both in carbonaceous and ordinary chondrites, with particularly forsteritic and refractory mineral chemistries and bright blue cathodoluminescence, which led him to entertain again a condensate origin (e.g. Steele (, Steele , for which the SIMS (Secondary Ion Mass Spectrometer) measurements of Weinbruch et al. (1993) found supporting evidence in their $^{16}$O enrichment (that is, in the direction of refractory inclusion compositions) and flat rare earth element (REE) patterns in olivine, inconsistent with igneous partitioning. Yet, as further data confirmed, the oxygen isotopic composition of the IOG largely overlap with chondrules in carbonaceous chondrites (e.g., Jones et al. (; Russell et al. (; Ushikubo et al. 2012). Also, *in situ* laser ablation inductively coupled mass spectrometer (LA-ICP-MS) analyses have revealed that IOG olivine has, in fact, quite fractionated REE, in accordance with near-equilibrium igneous partitioning (e.g. Jacquet et al. (; Jacquet and Marrocchi (; Pack et al. (2005. While the balance of evidence thus remains in favor of a genetic link between IOG and chondrules, some systematic differences, at least among the forsteritic ones, with the general population of chondrules remain to be understood so as to decide whether they sample the same heating events. For instance, the relative $^{16}$O enrichment of Tagish Lake (C2-ung) IOG indicated "more primitive material than the forsterite-rich chondrules" to Russell et al. (2010) while Jacquet and Marrocchi (2017) attempted to relate their coarse grain size and incompatible element enrichment of IOG in Northwest Africa (NWA) 5958 (C2-ung) to longer thermal processing. The IOG would be obviously important (olivine-rich) pieces of the puzzle of chondrule origin, if only their exact formation context *vis-à-vis* mainstream chondrules could be ascertained.

Our recent Secondary Ion Mass Spectrometer (SIMS)/electron microprobe (EMP) work on type I chondrules (that is, reduced, with $Fo_{90-100}$ olivine) with porphyritic olivine (PO) textures in carbonaceous chondrites has increased our understanding of the formation of olivine in these objects (Marrocchi et al. (, Marrocchi et al. We have identified Al-Ti-poor cores in central olivine grains whose oxygen isotopic deviation from the host revealed their relict nature and likely derivation from amoeboid olivine aggregate (AOA)-like precursors. These are overgrown by olivine enriched in incompatible elements and crystallized from melt produced during initial heating. Finally, palisadic olivine grains near the chondrule margins show evidence for gas-assisted outward growth, as also beautifully depicted by the cathodoluminescence (CL) maps of Libourel and Portail (2018). These different generations of olivine multiply our opportunities to find suitable analogs for IOG, at least for the magnesian population which dominates in carbonaceous chondrites. We have thus set to extend our work from type I chondrules to their isolated olivine counterparts (which we will also refer to as "type I IOG") in carbonaceous chondrites, where such reduced compositions dominate their high-temperature fraction. Specifically, we present in this paper combined petrographic (CL), chemical (EMP), oxygen isotopic (SIMS) data for magnesian IOG and chondrules in CM, CO and CV chondrites. The comparison will afford new insights on the genesis of these IOG.

## 2. Samples and methods

We surveyed the polished sections NWA 5958-1 and NWA 5958-4 of Northwest Africa (NWA) 5958, the thin section 3181lm4 (or 3181-4) of Allende (all three from the Muséum national d'Histoire naturelle, Paris, France), the thin section ALH 77307,96 of Allan Hills (ALH) 77307 (from the NASA Antarctic Search for Meteorites program) and a thick section of NWA 11086 prepared at CRPG. Allende is an oxidized Allende-like CV sub-group classified as a CV3.6 by Bonal et al. (2006). ALH 77307 is a CO3.0 chondrite (Grossman and Brearley 2005; Bonal et al. 2007; Busemann et al. 2007). Opaque assemblages, chondrule mesostases, olivine textures and compositions of chondrules show no indication of alteration and diffusional exchange. NWA 5958 corresponds to a C2-ung CM-like chondrite with type II chondrule olivine Cr content and opaque petrography indicating minimal thermal metamorphism (<

300°C) but chondrule mesostases bearing witness to significant aqueous alteration (Jacquet et al., 2016; Jacquet and Marrocchi 2017). NWA 11086 has been classified as a CM-an due to $^{16}$O enrichment and the absence of phyllosilicates according to X-ray diffraction (Gattacceca et al. 2019), but significant aqueous alteration is indicated by presence of PCP (Poorly Characterized Phases, also known as tochilinite cronstedtite intergrowths; see e.g. Lentfort et al., this issue), conversion of chondrule mesostasis to « spinach » (Fuchs et al. 1973) and paucity of Fe-Ni metal, so it might be a plain, if somewhat $^{16}$O-rich, CM2 chondrite. In the plots and the discussion, data from these latter two chondrites and ALH 77307 will be subsumed in the CM/CO clan.

Secondary Electron Microscope (SEM) imaging was performed on a JEOL JSM-6510 SEM equipped with a Genesis EDX detector at the Centre de Recherches Pétrographiques et Géochimiques (CRPG-CNRS, Nancy, France) using 3 nA electron beam accelerated at 15 kV. Modal abundances of high-temperature components were also estimated by manual point counting on BSE maps using the JMicrovision software (N. Roduit; https//jmicrovision.github.io; last accessed in December 2019) with 4000 points (randomly chosen by the software) per section (except for ALHA 77307, 2000 points). An object was considered an IOG if its interior was dominated by one olivine crystal. During closer BSE observation, only IOG large enough for SIMS analyses (~15 μm spot; this is also roughly the scale over which the BSE image allowed assignment to IOG in the point counting) were selected for further study. The effective radius of the IOG or the coarsest olivine of each studied chondrule was calculated as the radius of the equal area disk, from back-scattered electron (BSE) images. Cathodoluminescence (CL) imaging of chondrules was performed using (i) an RGB CL-detection unit attached to a field emission gun secondary electron microscope JEOL J7600F at the Service Commun de Microscopie Électronique (SCMEM, Nancy, France) and (ii) a field-gun ZEISS Supra 55 VP equipped with an OPEA catholuminescence device (imaging and spectral) with a high-tech parabolic mirror. Quantitative chemical compositions of olivine grains were obtained using a Cameca SX Five electron microprobe (EMP) at the Université Pierre et Marie Curie (UPMC, Camparis, Paris, France) using a 150 nA focused beam (about ~2 μm in diameter) accelerated to 15 kV. We analysed Na, Mg, Si, Al, K, Ca, Fe, Ti, Cr, and Mn in olivine grains. The high beam current allowed detection limits for silicates of 100 ppm for Al, Ca, and Ti, 150 ppm for Mn and Si, and 200 ppm for Na, K, Cr, Fe, and Mg. The PAP software was used for matrix corrections.

We measured the oxygen isotopic compositions of chondrule olivine and isolated olivine grains, where CL and EMP had shown fairly homogeneous compositions (typically near the center), with a CAMECA ims 1270 E7 at CRPG-CNRS (Nancy, France). $^{16}O^-$, $^{17}O^-$, and $^{18}O^-$ ions produced by a Cs$^+$ primary ion beam (~15 μm, ~4 nA) were measured in multi-collection mode with two off-axis Faraday cups (FC) for $^{16,18}O^-$ and the axial FC for $^{17}O^-$. The FC had $10^{11}$ Ω amplifiers. To remove $^{16}OH^-$ interference on the $^{17}O^-$ peak and to maximize flatness atop the $^{16}O^-$ and $^{18}O^-$ peaks, the entrance and exit slits of the central FC were adjusted to obtain mass resolution power of ~7000 for $^{17}O^-$. As an additional safeguard against $^{16}OH^-$ interference, a N$_2$ trap was used to reduce the pressure in the analysis chamber to <5 × 10$^{-9}$ mbar. The multicollection FCs were set on exit slit 1 (MRP = 2500). Total measurement times were 240 s (180 s measurement + 60 s pre-sputtering). We used three terrestrial standard materials (San Carlos olivine, magnetite and diopside) to define the instrumental mass fractionation line for the three oxygen isotopes and correct for instrumental mass fractionation for olivine. To obtain good precision on analytical measurements, we analyzed, in order, 4 standards, 8 chondrule olivine crystals and 4 standards. Typical count rates obtained on the San Carlos olivine standards were 2.5 × 10$^9$ cps for $^{16}O$, 1.0 × 10$^6$ cps for $^{17}O$, and 5.4× 10$^6$ cps for $^{18}O$. The isotopic compositions are expressed in standard δ-notation, relative to Vienna standard mean ocean water (VSMOW): $\delta^{18}O = (^{18}O/^{16}O)_{sample}/(^{18}O/^{16}O)_{VSMOW} - 1$ and $\delta^{17}O = (^{17}O/^{16}O)_{sample}/(^{17}O/^{16}O)_{VSMOW} - 1$ (both to be expressed in ‰). 2σ measurement errors, accounting for internal errors on each measurement and the external reproducibility of the standard, were estimated to be <1‰ for $\delta^{18}O$, $\delta^{17}O$ and $\Delta^{17}O \equiv \delta^{17}O - 0.52 \times \delta^{18}O$ (representing deviation from the Terrestrial Fractionation Line = TFL).

## 3. Results

Petrography

Type I chondrules present the usual mineralogical zoning (Fig. 1) of olivine phenocrysts dominating in the interior and enstatite laths, poikilitically enclosing olivine chadacrysts, near the outside (e.g., Friend et al. (; Libourel et al. (2006. The mesostasis is altered in the CM chondrites. When viewed in CL, interior olivine grains often display dark cores (Fig. 1) that correspond to relict grains for those chondrules investigated by Marrocchi et al. (2018, 2019), surrounded by brighter overgrowths (identified to the *in situ* crystallized host). The "palisadic" olivine grains near the outside of PO chondrules (Marrocchi et al. 2018, 2019; Libourel and Portail 2018) have brightest CL in their inner edge and darken toward the exterior (Fig. 1D).

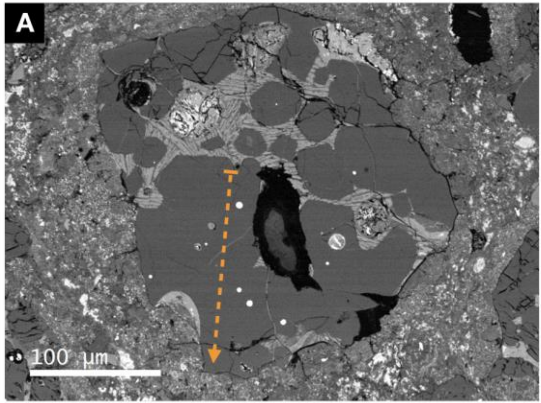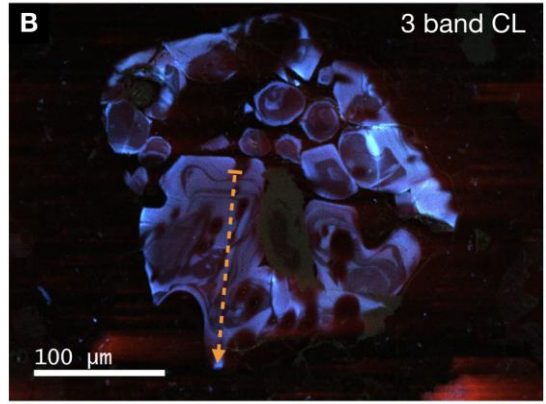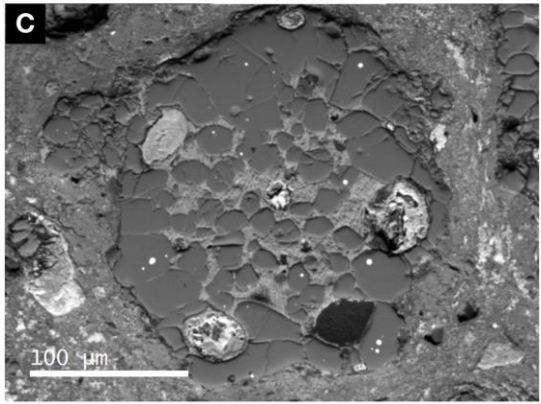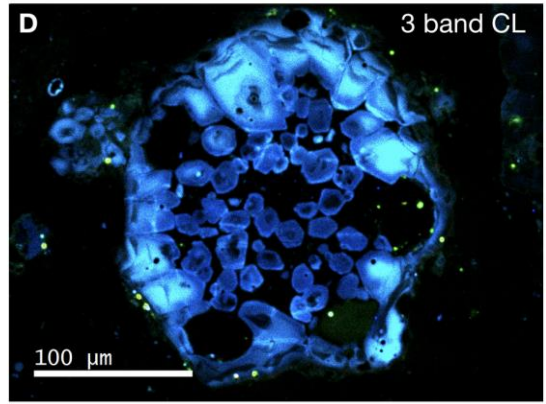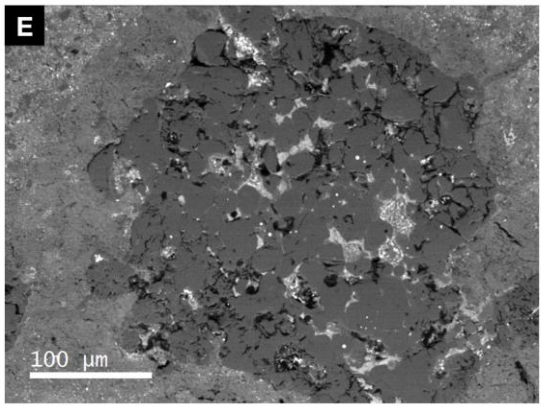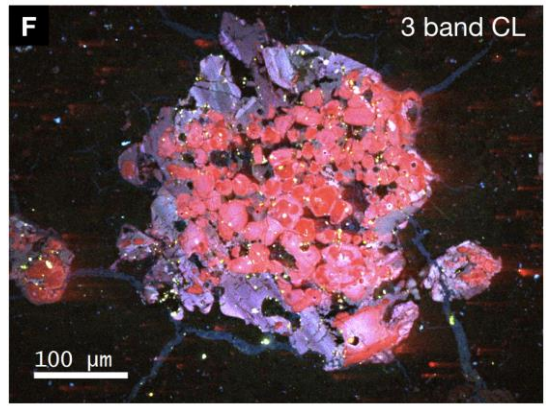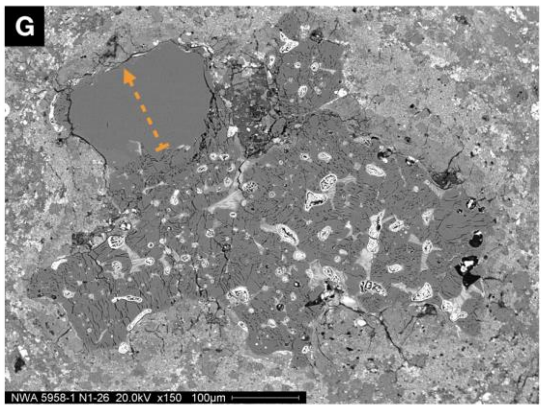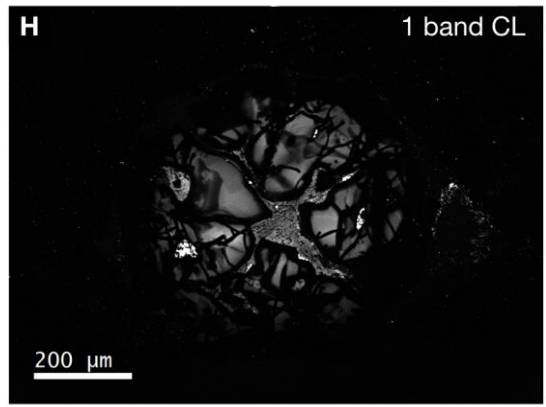

**Figure 1**: Back-scattered electron and cathodoluminescence images of representative chondrules (left and right, respectively, unless otherwise noted). Orange dashed lines show location of profiles shown in Fig. 6. A) and B) Chondrule CHA in section NWA 5958-4. The ovoid chondrule contains euhedral olivine grains with little discernible pyroxene and relatively abundant altered mesostasis. Its lower half is occupied by a large olivine apparently embayed near the central hole, and has concentric and oscillatory zoning in blue CL (as the other olivine phenocrysts). Many of the oval dark-CL spots trace metal inclusions. C) and D) PO chondrule CH9 in section NWA 5958-1. Dark-CL spots in interior olivine grains likely correspond to relicts. Brightest blue CL is seen in the coarser palisadic olivine grains but CL intensity declines with some oscillations near the outer edge. A small (~50 μm diameter) PO chondrule is attached to the upper left of the object. E) and F) POP chondrule CHA in NWA 11086. Olivine grains luminesce in red in the chondrule interior, with those closest to the center exhibiting large darker-CL (presumably relict) cores. In the periphery, enstatite oikocrysts have only weak, bluish CL (note that left panel is somewhat more zoomed than right panel). G) Chondrule N1-26 in NWA 5958 (BSE only). This irregularly shaped chondrule is dominated by pyroxene crystals, with <20 μm olivine grains, in addition to altered mesostasis and metal grains. However, one large (0.2 mm across) olivine crystal (analyzed by Jacquet and Marrocchi 2017) is visible on the upper left edge of the chondrule, partly surrounded by a thin layer of enstatite and must have been an IOG fused with the chondrule while still plastic (otherwise it would have been replaced by pyroxene to the same extent as the remainder of the chondrule margins). H) Chondrule CH1 in Allende (CL only) shows coarse, bright CL triangular olivine crystals and mesostasis near the center. It might be a surficial section through the "palisade" of a PO chondrule.

**Table 1**: Modal abundances of high-temperature components in the studied chondrites.

| Meteorite | NWA 5958 | NWA 11086 | ALHA 77307 | Allende |
|---|---|---|---|---|
| Classification | C2-ung | CM-an | CO3.0 | CV3 |
| *Type I chondrules* | | | | |
| Bona fide | 20.3 | 23.2 | 36.3 | 31.3 |
| IOG | 2.3 | 3.0 | 2.6 | 1.5 |
| *Type II chondrules* | | | | |
| Bona fide | 1.2 | 1.4 | 3.8 | 1.0 |
| IOG | 0.9 | 1.1 | 1.6 | 0.2 |
| *Refractory inclusions* | | | | |
| AOA | 1.2 | 0.8 | 1.6 | 5.7 |
| CAI | 1.0 | 0.5 | 0.6 | 3.8 |

Point counting reveals 1.7-4.2 vol% of IOG in the studied carbonaceous chondrites (Table 1). Fuchs et al. (1973) reported 8 vol% of isolated mineral grains in Murchison, about two thirds of which appeared to be olivine; Browning et al. (1996) reported 6 vol% of IOG in Murchison, and down to 1.5 vol% in more altered CM chondrites. The average 22 vol% of "single grain and grain fragments" of Grossman and Olsen (1974) for C2 chondrites must be in error as it equals the *total* high-temperature fraction normally seen in CM chondrites (e.g., Table 1; Howard et al. (2011). As for CO3 chondrites, McSween (1977) quoted 8 vol% for IOG. While our numbers thus tend to be systematically lower than past literature estimates based on optical microscopy, we note that Pack et al. (2004) never found more than 0.35 vol% of "refractory forsterites" in carbonaceous chondrites. Since their refractory forsterites (recognized by CL) had CaO > 0.4 wt% and such compositions represent $2/5^{th}$ of our IOG analyses, and since Pack et al. (2004) only counted the luminescing part of their olivine, this would indicate IOG modal fractions of order 1 vol%. More recently, manual identifications on X-ray maps by Ebel et al. (2016) resulted in an average of 1.16 vol% of "isolated olivine grains or aggregates in matrix" in CO chondrites and 1.37 vol% in Allende. Part of the differences between studies may lie in the ambiguity in discriminating IOG from chondrules. Despite this systematic uncertainty, an order of magnitude of a few percent for IOG modes in carbonaceous chondrites seems overall sound.

While most IOG are of type I, it is noteworthy that the type II/type I modal ratio is significantly higher for IOG (0.1-0.6) than for *bona fide* chondrules (0.03-0.11), by a factor of 4 to 7. This is qualitatively in line with the histograms of McSween (1977) which indicate values of 2 and 0.8 for this ratio in CM and CO chondrites, respectively, with that of Desnoyers (1980) indicating a ratio of 1.2 for Niger (C2), although those histograms are based on microprobe analyses and not areas (perhaps accounting for the systematic difference). Seven out of the 12 Tagish Lake IOG of Russell et al. (2010) were of type II; same for 25 out of 101

CI chondrite IOG compiled by Piralla et al. (2020). This point being made, we henceforth exclusively focus on type I IOG.

As noted by Olsen and Grossman (1978), the IOG tend to be big compared to chondrule phenocrysts (Fig. 2, even though the first bin is cut off by our selection biases). In fact, in CV chondrites, their size distribution is comparable to that of the *coarsest* chondrule phenocrysts; in CM/CO chondrites, more than 90 % of the IOG studied here are even bigger than the mode of the coarsest chondrule phenocrysts (near 30 µm effective radius). About half (21 out of 41[1]) of the IOG examined here by BSE/CL and EMP have equidimensional, often euhedral/subhedral shapes, although the others are evidently fragments of larger objects (e.g. Figs. 3C,D; 4G,H). The CL of the former, generally blue in the core, is concentrically zoned, with intensity decreasing toward the margin which may have a redder tint. Discrete darker-CL streaks, partially or entirely surrounding the center, are often superimposed on the background trend, generally paralleling the grain edges. An oscillatory sector zoning is spectacularly visible in NWA 11086 IO12 (Fig. 5); some may be discernible in NWA 5958-1 IO7 as well (Fig. 4F). Despite their name, the IOG are not pure olivine. The olivine may enclose metal grains and glass inclusions, and, when whole, is nearly always surrounded, partly or entirely, by enstatite (from a few microns to tens of microns in thickness), sometimes with mesostasis (although we did not find as large a mesostasis patch as in IOG N6-5 in Fig. 1e of Jacquet and Marrocchi 2017) and even further olivine phenocrysts. Enstatite rims, partial or total, are sometimes seen even around fragments (e.g. Fig. 3C,D). We note that Jones (1992) called large olivine grains surrounded by pyroxene in ALHA 77307 "macroporphyritic chondrules", in contradistinction to "bare" isolated olivine grain, but given the continuum between them, we maintain the name IOG for all those objects. The continuum, in fact, extends to *bona fide* chondrules (e.g. Jacquet and Marrocchi 2017), with some chondrules exhibiting disproportionately large olivine crystals (e.g. Fig. 1A,B; chondrule N5-21 with three aligned coarse olivine grains in Fig. 2b of Jacquet et al. 2016), not to mention chondrule-IOG or IOG-IOG compounds (see e.g. Fig. 1G, 3E,F). Overall, this continuum makes the assignment of objects to IOG rather than chondrules relatively subjective and must have contributed to the scatter in IOG modes in the literature discussed above. We recall that we have considered here an object to be an IOG when its interior was dominated by one olivine crystal.

---

[1] Fragments may be numerically more numerous for smaller sizes which we have not selected. Still, it may not affect the modal ratio much (here the studied "whole" IOG outweigh fragments also in this respect, with a total area of 0.74 mm² vs. 0.59 mm²).

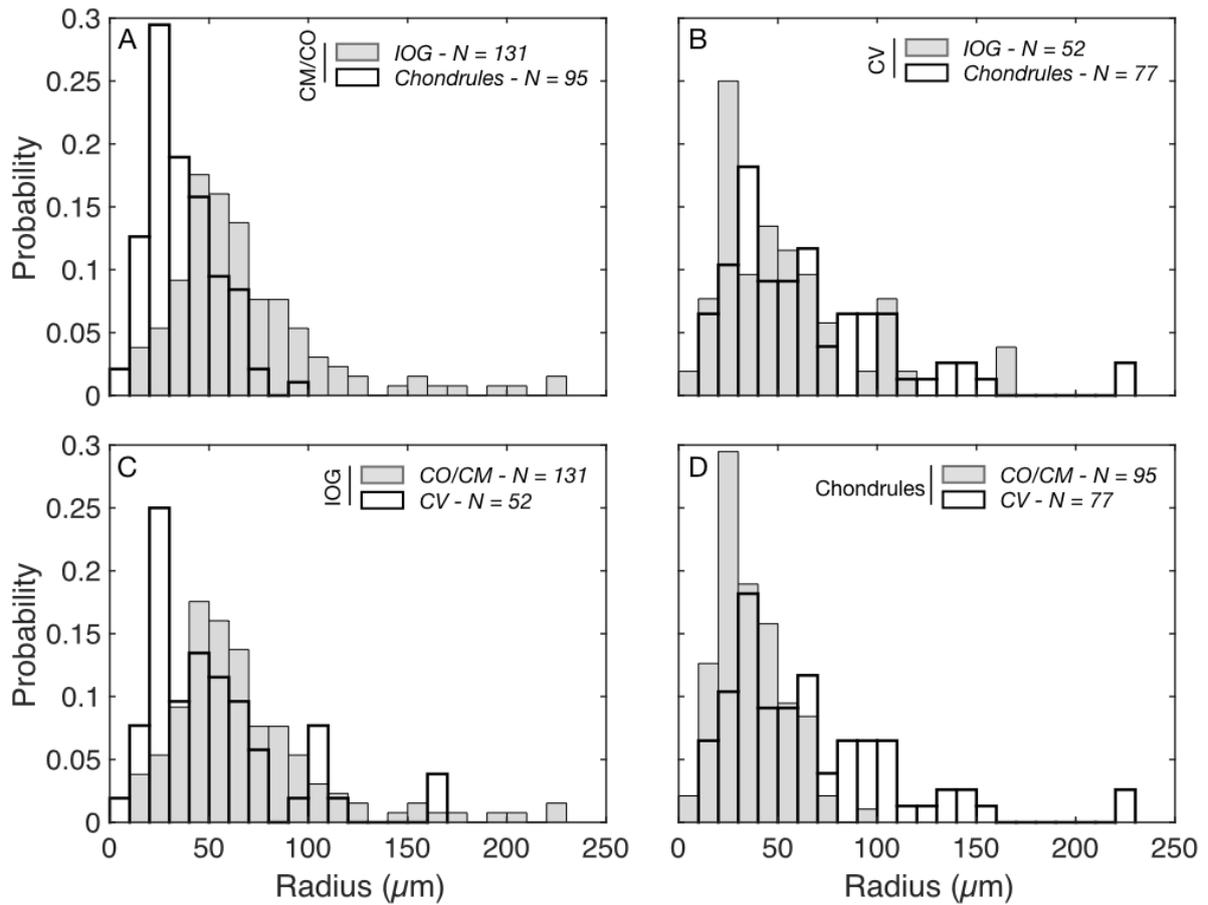

**Figure 2**: Histograms of sizes of IOG and coarsest olivine grains of chondrules. A) and B) show data for CO/CM and CV chondrites, respectively, for intra-clan comparison, while C) and D) allow inter-clan comparison (for IOG and chondrules, respectively).

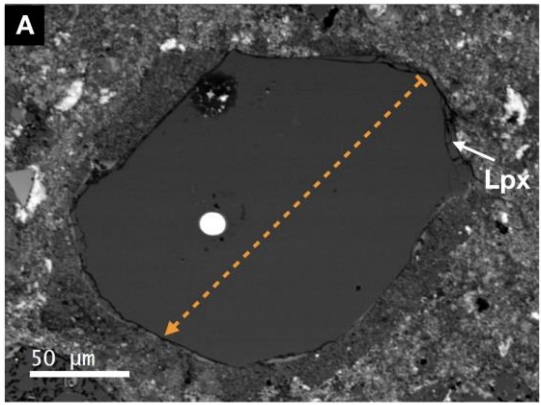
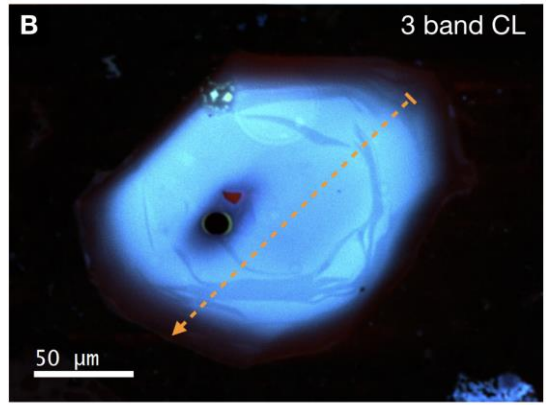
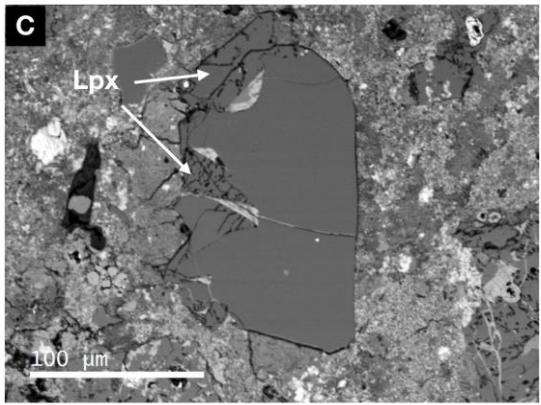
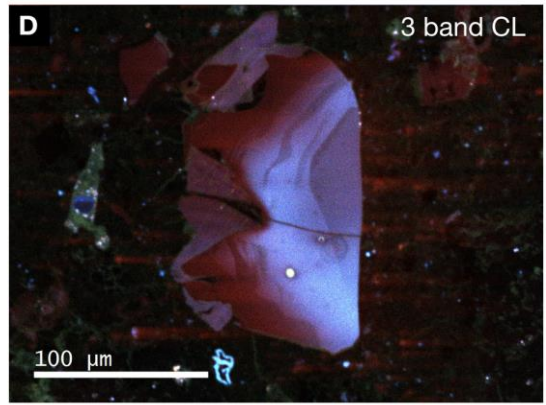
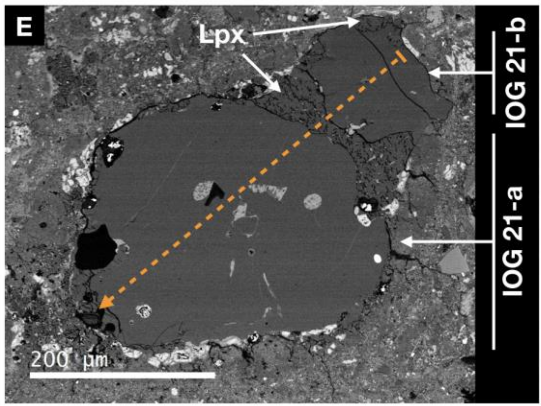
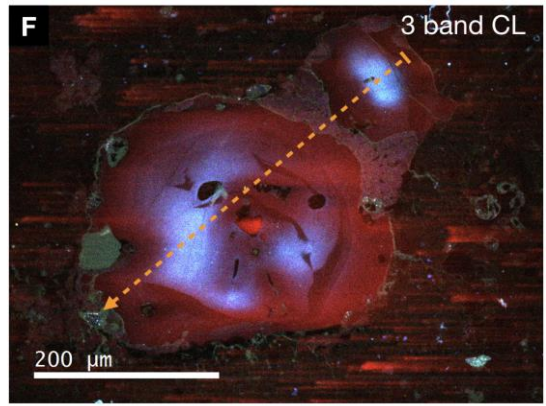
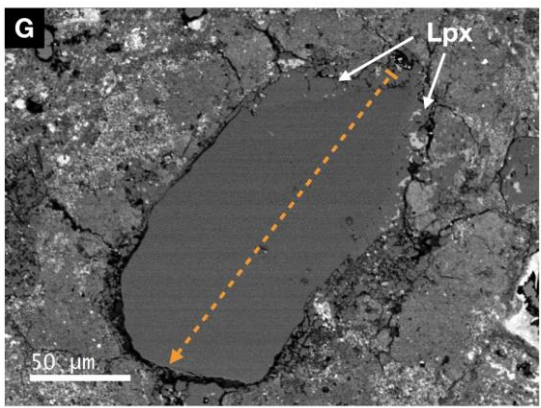
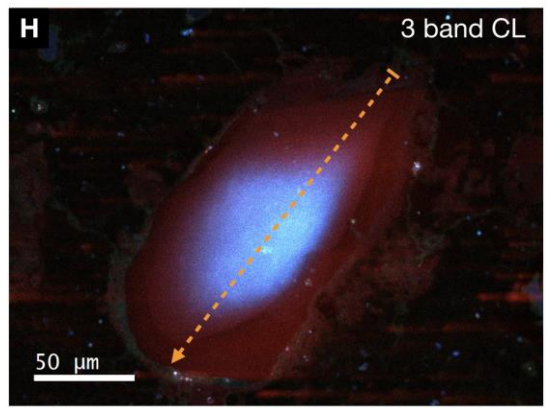

**Figure 3**: Back-scattered electron (left) and cathodoluminescence (right) images of representative isolated olivine grains. "Lpx" indicates enstatite and orange dashed lines locate the profiles shown in Fig. 6. A) and B) NWA 5958-4 IO5. This subhedral grain has a thin (<5 µm) enstatite layer. The olivine has bright blue CL over its interior except its outermost reddish/black-CL 10 µm. Somewhat darker blue-CL streaks occur concentrically. A metal grain is surrounded by a dark-CL ellipsoidal halo, perhaps due to anisotropic diffusion of $Fe^{2+}$ therefrom. C) and D) NWA 5958-4 IO13. This IOG presents a ~20 µm thick pyroxene + (altered) mesostasis layer on the left, with an irregular boundary with the olivine. The latter seems abruptly cut off on the right side. CL is zoned from right (blue) to left (reddish), and would have been roughly concentric around a center located outside the present right edge. E) and F) NWA 5958-4 IO21 is a compound between two IOG, tied by a pyroxene neck which extends to thin layers around the peripheries of the two olivine grains. The compound must have formed by collision between two IOG which had independently acquired a pyroxene margin, as otherwise the olivine would have been continuous across the two lobes. The CL is zoned from bright blue cores to reddish margins. The zoning of the larger IOG (NWA 5958-4 IO21a) is complicated by dark-CL streaks and altered mesostasis inclusions, one with spinel. G) and H) NWA 5958-4 IO36. This elongated euhedral olivine grain is CL-zoned (from bright blue to red) parallel to the edge and is surrounded by enstatite, thickest around the upper acute angle.

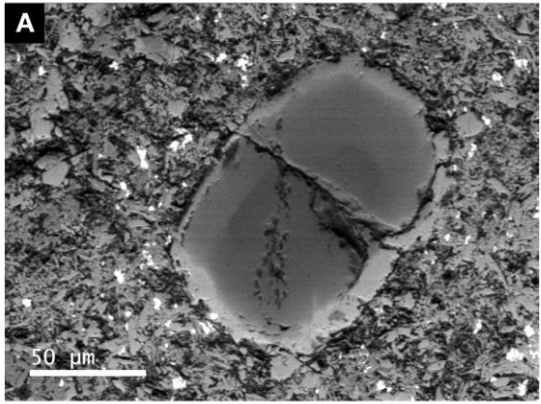 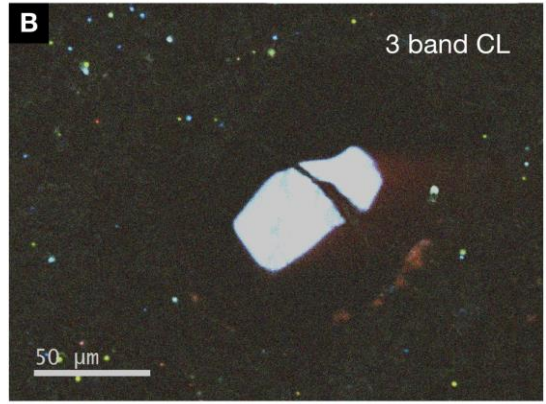
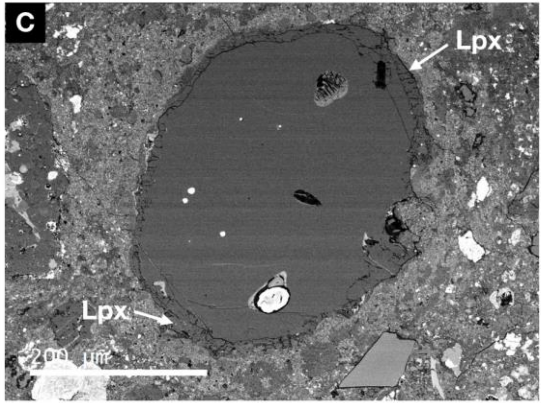 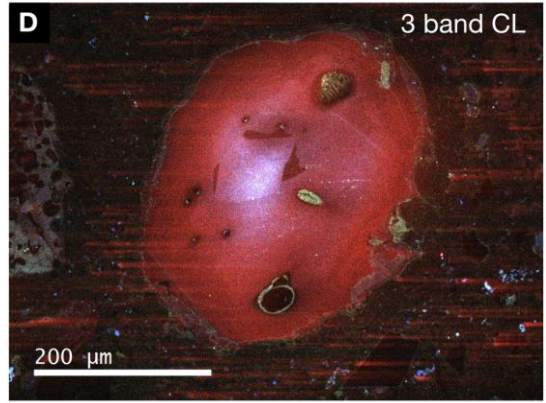
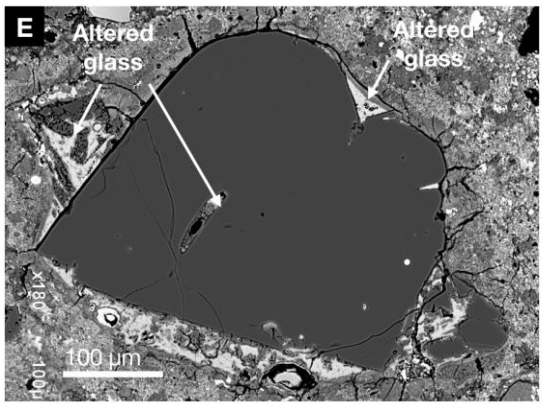 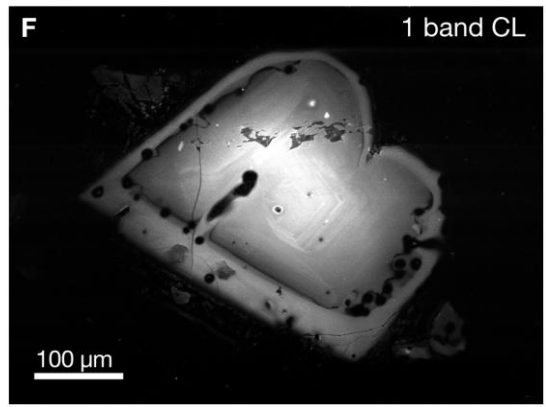
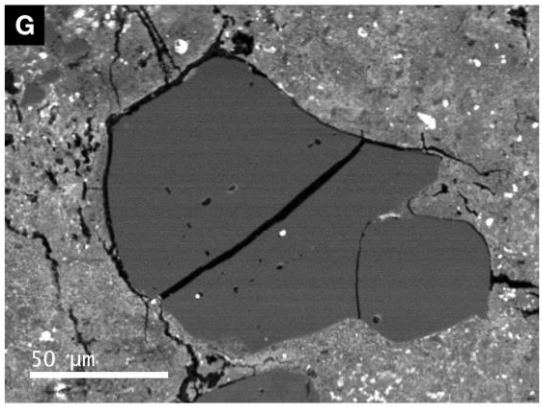 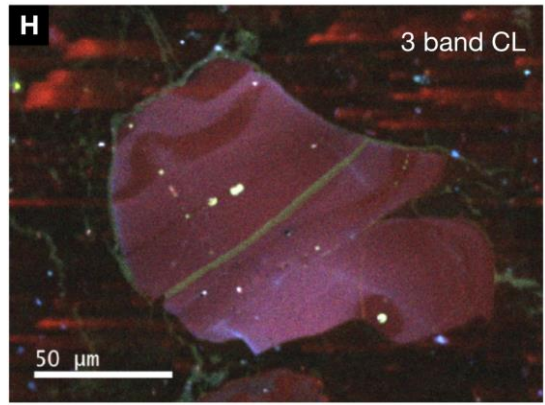

**Figure 4**: Back-scattered electron (left) and cathodoluminescence (right) images of additional representative isolated olivine grains. "Lpx" indicates enstatite. A) and B) Allende IO19. This elongated subhedral grain shows a bright blue CL in its forsteritic core, mantled without transition by more ferroan (dark-CL) olivine itself surrounded by a BSE-bright more fayalitic outer layer betraying influx of Fe during metamorphism. C) and D) NWA 5958-4 IO22. This oval olivine grain shows red CL, brightest near the center, and is surrounded by a continuous 10 µm-thick layer of enstatite (darker than olivine in BSE, pink in CL). Altered glass inclusions and one big altered and small, unaltered, metal grains also occur inside the olivine. E) and F) NWA 5958-1 IO21 (the CL image is here a one-band panchromatic image so as to leave the details discernible). This subhedral olivine grain has a bright-CL margin separated from a core with oscillatory zoning by a boundary strikingly parallel to the edge of the whole grain. The olivine is embedded in altered mesostasis (with pyroxene phenocrysts), with two discrete patches on the left and the right (the latter with olivine crystals), possibly attached during cooling. G) and H) NWA 5958-4 IO16. This is clearly a fragment. The darker-CL streaks in this red-CL object show no concentric arrangement.

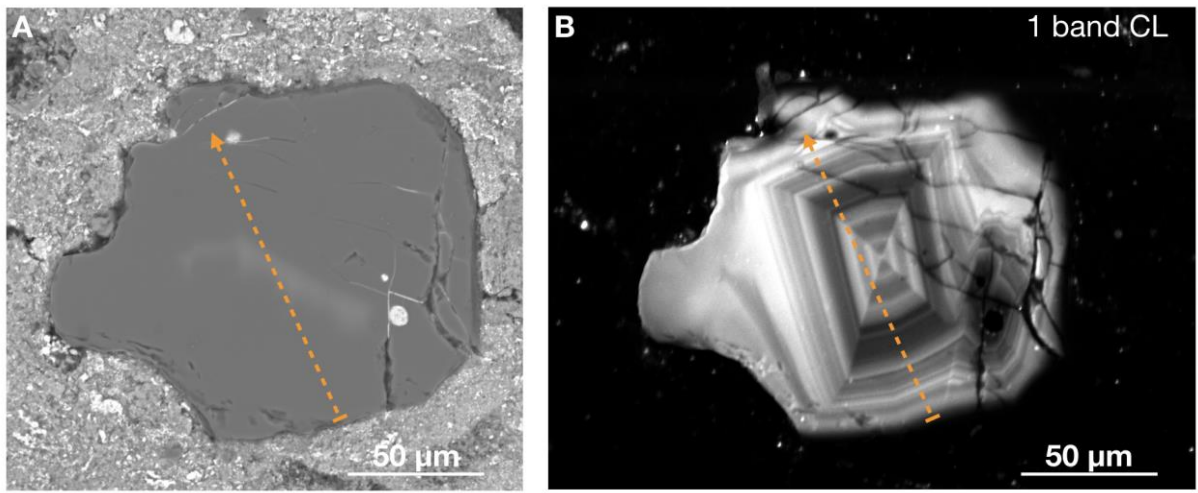

**Figure 5**: Back-scattered electron (A) and cathodoluminescence (B) images of isolated olivine grain NWA 11086 IO12, with dashed line indicating the location of the profile shown in Fig. 6H. The CL shows spectacular oscillatory sector zoning in a central rectangular area. On the lower right, a darker-CL irregularly shaped angular patch seems to indicate an independent olivine grain (enclosing two black-CL metal grains) welded together with the previous one, with subsequent layers with oscillatory zoning swathing both. The irregular outlines may point to fragmentation around the core of the parent olivine.

### Chemistry and oxygen isotopes

Individual profile location and analyses are shown in the Electronic Annex and some representative ones are illustrated in Fig. 6. The CL zoning in IOG olivine is reflected by EMPA traverses, with Ca, Al, Ti decreasing from core to rim, whereas Fe, Mn, Cr symmetrically increase outward. Thus the zoning is basically of decreasing "refractoriness" toward the margin (Fig. 6; Jones 1992; Jacquet and Marrocchi 2017). This is seen in the coarsest chondrule olivine grains as well (more asymmetrically for the palisadic grains, with 0.2-0.5 wt% CaO; Marrocchi et al. 2018) although many have flatter minor element profiles (Fig. 6A-D). In general, excluding obvious fragments, flat profiles are associated with uniformly low CaO (<0.4 wt%) abundances. While Al and Ti roughly parallel Ca, they may suffer abrupt drops or peaks (e.g. Fig. 6A,D) which correspond to oscillations seen in CL (whose lengthscales may be shorter than the profile steps) or occasional relict grains (mostly for chondrules; Marrocchi et al. 2018). This is consistent with the contention of Libourel and Portail (2018) that CL intensity is largely controlled by Al (in the absence of sufficient amounts of the CL quencher Fe). Only in NWA 11086 IO12 (Fig. 6H) does Ca show oscillations unattenuated relative to Al and Ti.

In the biplots of this paper (viz. Fig. 7, 8), the IOG will be represented by their apparent core composition. CaO spans 0.16-0.89 wt% in IOG and 0.05-0.66 wt% in chondrule coarsest olivine (with averages of 0.44 vs. 0.30 wt%); $Al_2O_3$ spans 0.06-0.42 wt% in IOG and chondrule coarsest olivine 0.02-0.53 wt% (with averages of 0.20 vs. 0.12 wt%). Crystal size seems to correlate negatively with Fe, Mn but the positive correlation with Ca and Al noted by Jacquet and Marrocchi (2017) is very rough (Fig. 7). This may be to some extent a 2D sectioning artifact as more equatorial sectioning of the grains (providing the widest areas) would pass closest to the actual refractory core. Excluding Allende, whose olivine FeO contents up to 5 wt% are obviously secondary, the IOG have low FeO contents (0.35-1.4 wt%) anticorrelated with CaO, similar to chondrule phenocrysts in this compositional range (Fig. 8).

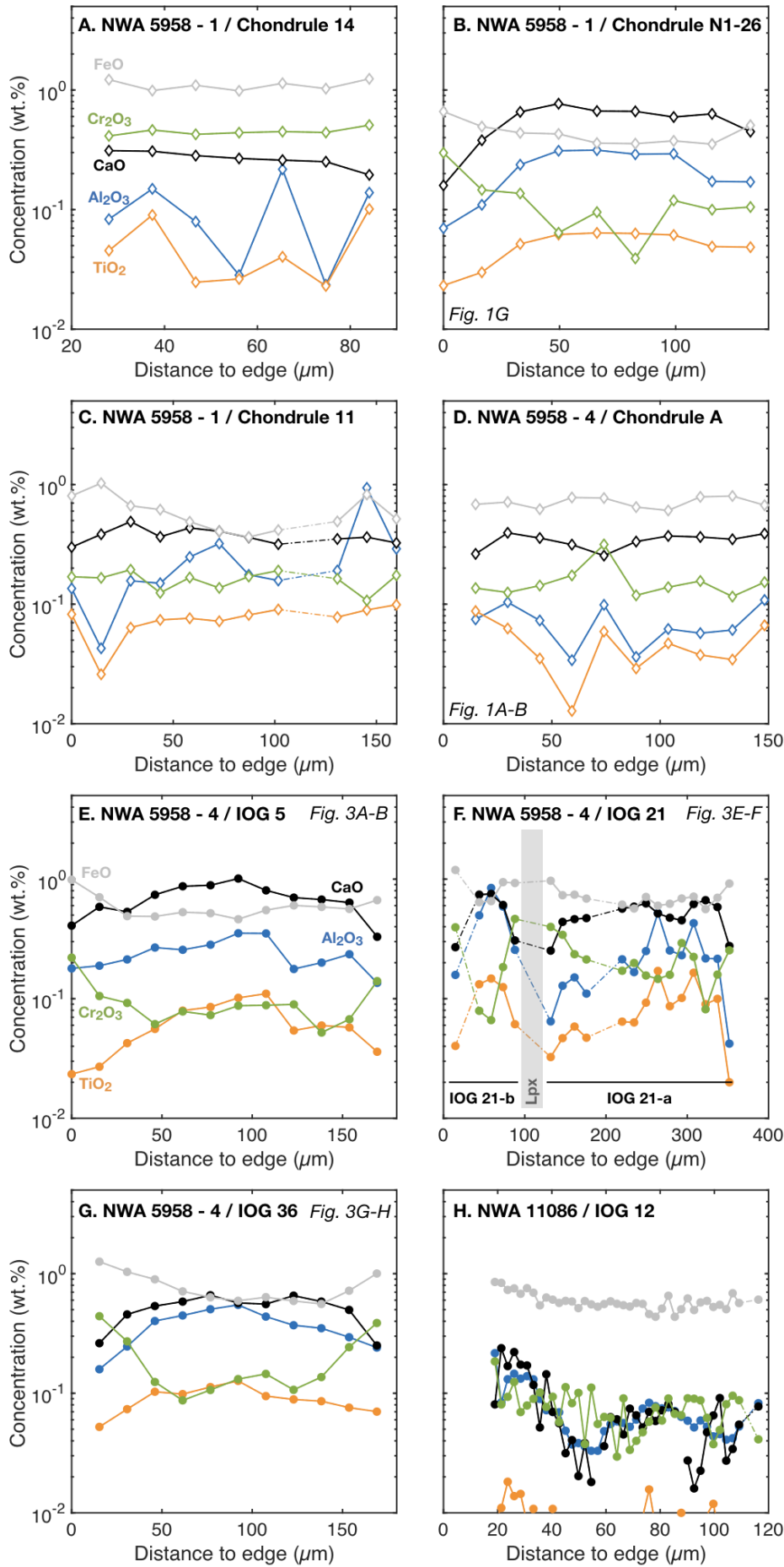

**Figure 6:** Minor element profiles in chondrule olivine (A-D) and IOG (E-H, with profile positions in the indicated figures). Recall that NWA 5958-4 IOG21 is a compound object where the profile transects two successive (formerly isolated) olivine grains (Fig. 3E,F).

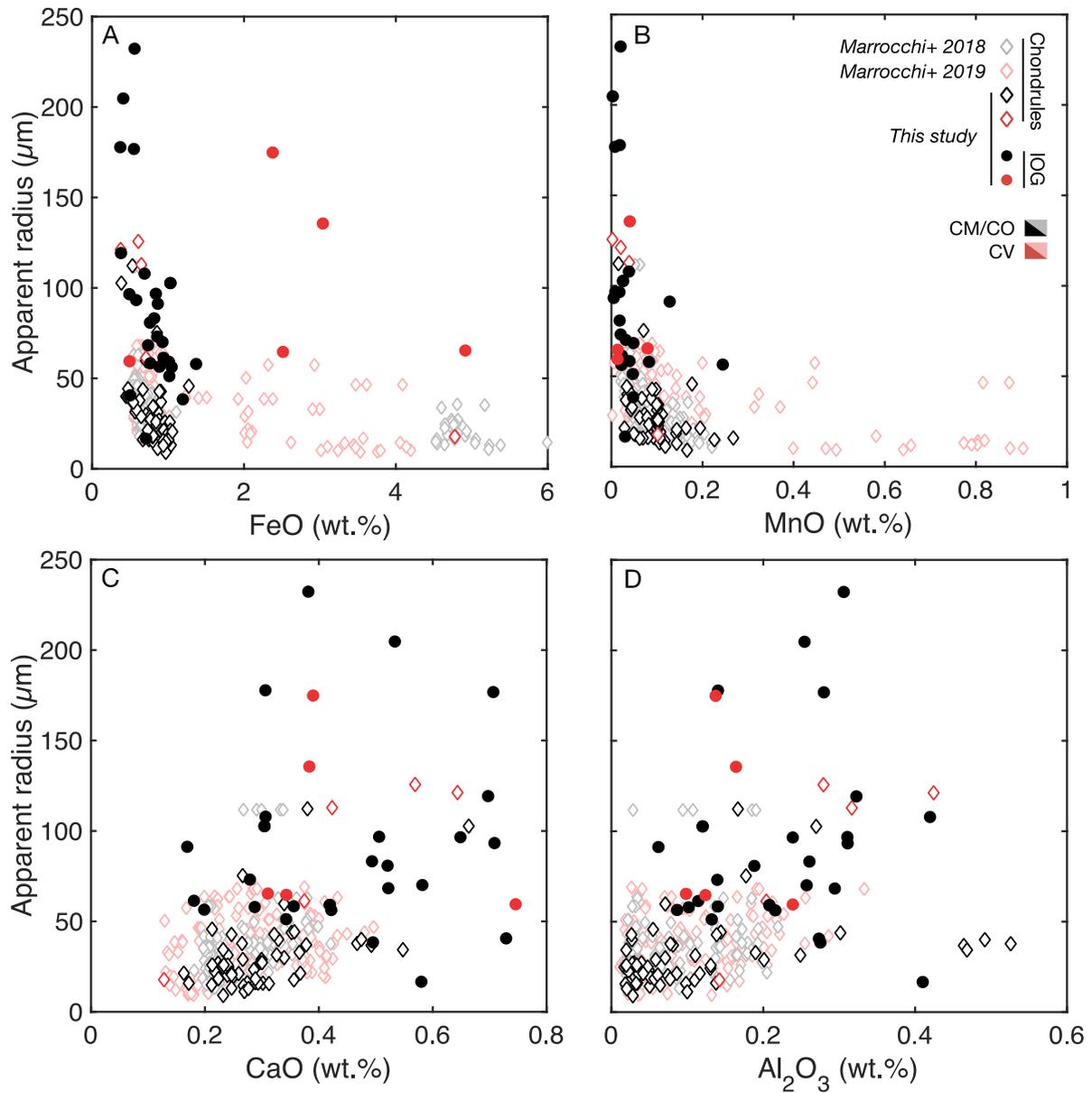

**Figure 7**: Olivine size vs minor element content (A: FeO; B: MnO; C: CaO; D: $Al_2O_3$) in IOG and chondrule coarsest olivine. Data for all analyzed chondrule phenocrysts in Marrocchi et al. (2018, 2019) are also shown.

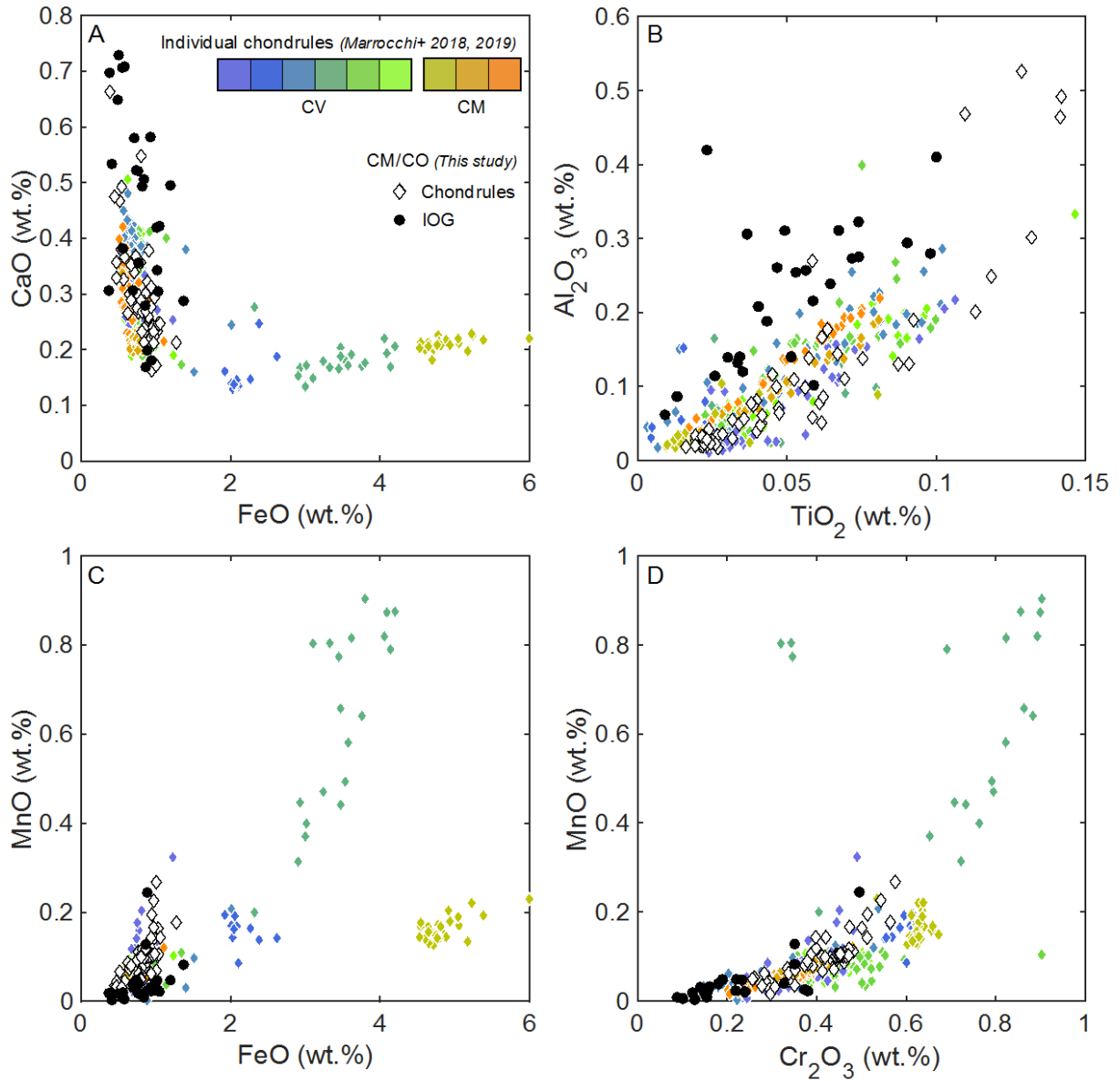

**Figure 8**: Minor element biplots of IOG, chondrule coarsest olivine of this study and chondrule olivine (keyed by chondrule) from Marrocchi et al. (2018, 2019). A) CaO vs. FeO. The correlation is negative for FeO below ~1.5 wt% and positive afterward. Individual chondrules plot in only one of these trends. IOG all belong to the former. B) $Al_2O_3$ vs. $TiO_2$. C) MnO vs. FeO. D) MnO vs $Cr_2O_3$.

The oxygen isotopic compositions of chondrule olivine and IOG plot along the Primitive Chondrule Mineral (PCM; Fig. 9) line of Ushikubo et al. (2012). When analyzed multiple times, the IOG appear generally isotopically homogeneous, with the exception of NWA 11086 IO22 (with $\Delta^{17}O$ of -0.18 ± 0.60 ‰ and -5.39 ± 0.60 ‰), but the number of analyses per object is usually small. $\Delta^{17}O$ ranges from -12.65 to -0.2 ‰ for chondrules and -7.46 to 0.84 ‰ for IOG. The distribution (Fig. 10) is fairly similar to the chondrule host (i.e., non-relict) olivine grains of Marrocchi et al. (2019), with a minor $^{16}O$-poor population above the ~ -4 ‰ hiatus noted by Ushikubo et al. (2012). The four IOG in this range are not more fayalitic than the others, unlike the trend shown by type I chondrules in several studies (e.g. Ushikubo et al. 2012; Schrader et al. 2013; Tenner et al. 2013, 2015) but similar to the Jacquet and Marrocchi (2017; see their Fig. 8a) data for NWA 5958. Nevertheless, high refractory element contents (e.g. CaO contents above 0.4 wt%)—corresponding to the refractory forsterites of Steele (1988)—are restricted to $^{16}O$-rich IOG or *bona fide* chondrules with $\Delta^{17}O < -4$ ‰ (Fig. 11; Libourel and Chaussidon 2011; Jacquet and Marrocchi 2017; Marrocchi et al. 2019).

The isotopic, chemical and geometrical data for all analyzed objects are shown in the Electronic Annex.

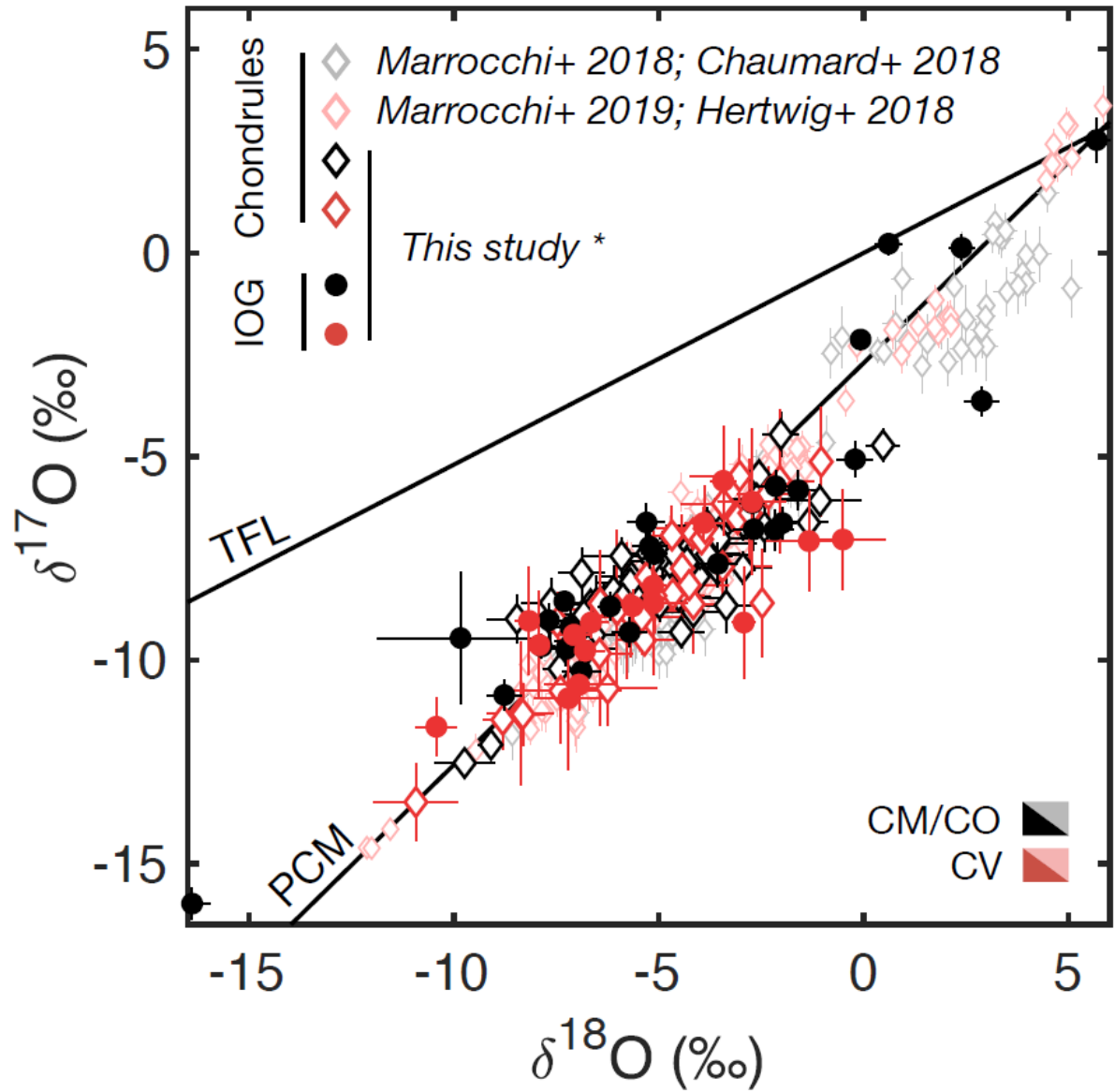

**Figure 9**: Three-isotope plot of olivine grains in this study, including CV data of Libourel and Chaussidon (2011), with chondrule literature data from Marrocchi et al (2018, 2019), Chaumard et al. (2018), Hertwig et al. (2018). The Primitive Chondrule Mineral (PCM) line of Ushikubo et al. (2012) is also shown,

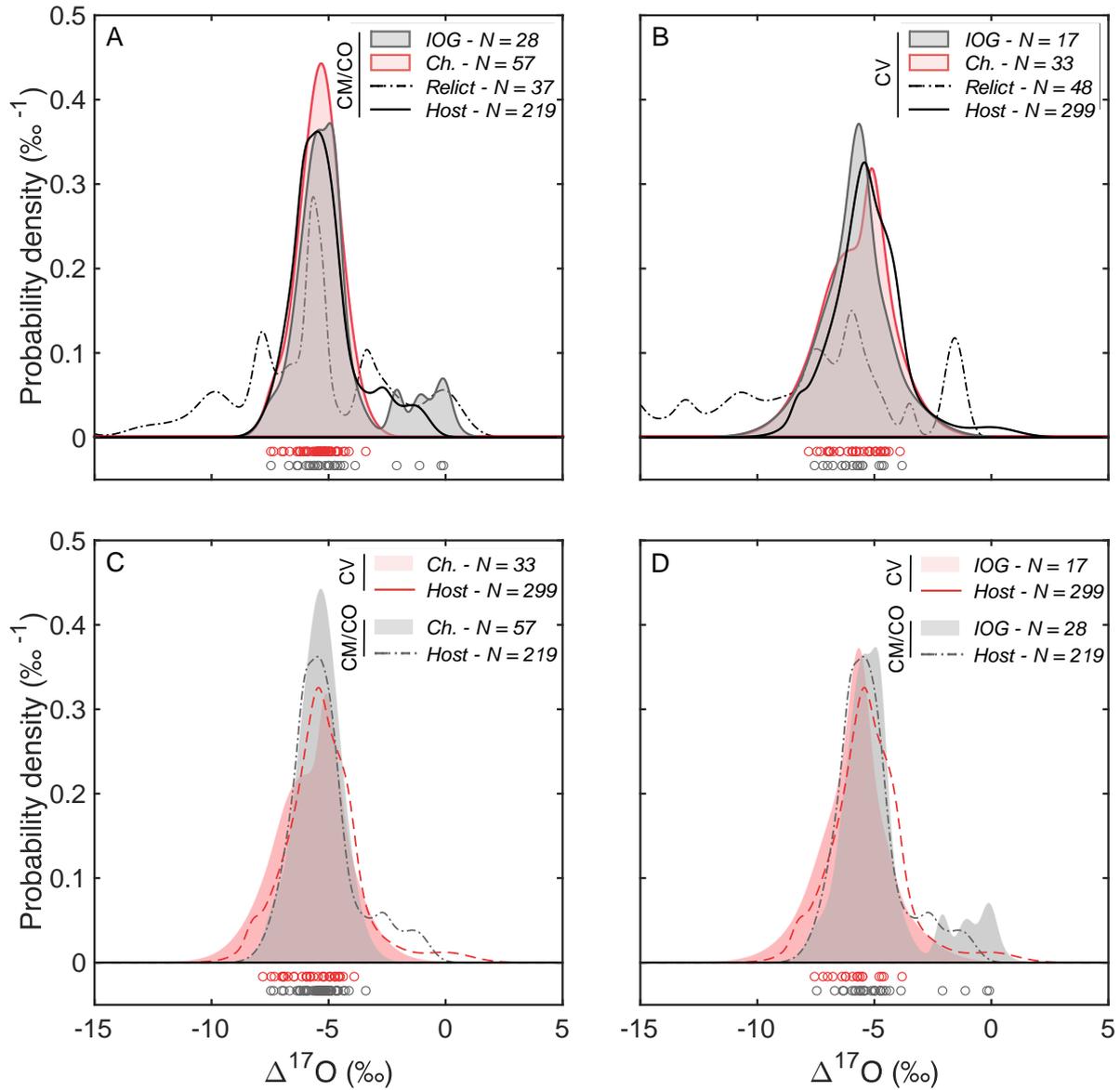

**Figure 10**: Probability density function of $\Delta^{17}O$ for IOG and coarsest chondrule olivine, compared to host and relict data compiled by Marrocchi et al. (2019). A) and B) compare data within chondrite clans (CM/CO and CV respectively) and C) and D) compare chondrule and IOG data across the clans.

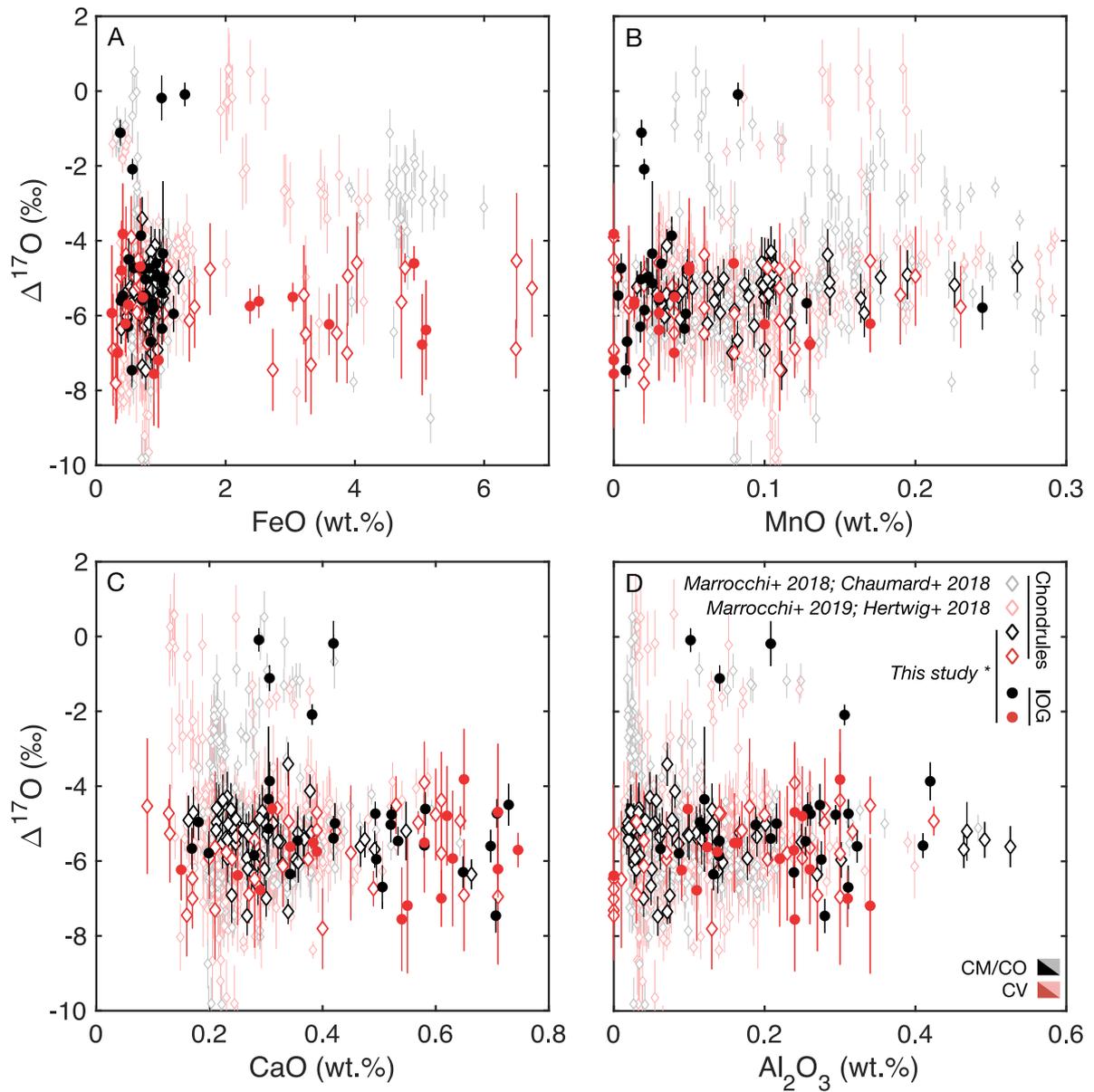

**Figure 11**: Oxygen isotopes vs. minor elements for IOG and coarsest olivine grains, including CV data by Libourel and Chaussidon (2011), along with host olivine data from the Marrocchi et al. (2019) compilation. A) $\Delta^{17}O$ vs. FeO. B) $\Delta^{17}O$ vs. MnO. C) $\Delta^{17}O$ vs. CaO. D) $\Delta^{17}O$ vs. $Al_2O_3$.

# 4. Discussion

4.1 Genetic link between isolated olivine grains and chondrules

The general overlap in $\Delta^{17}O$ (Fig. 10) between IOG and chondrule host olivine, as well as in minor elements, indicates, as in previous work (McSween 1977; Desnoyers 1980; Jones 1992; Jones et al. 2000), that IOG are closely linked to chondrules rather than early nebular condensates such as AOAs. The presence of mesostasis is also evidence for an igneous origin incompatible with gas-solid condensation (e.g. McSween 1977; Richardson and McSween 1978). Other phases associated with olivine (metal, occasional companion olivine crystals, pyroxene) are also most analogous to chondrule petrography; with IOG CL being comparable to that of the coarsest olivine in chondrules (Fig. 1). We may also recall the REE fractionation measured in IOG olivine (Pack et al. 2005; Jacquet et al. 2012; Jacquet and Marrocchi 2017) which is at variance with the flatter patterns in AOA olivine which formed by condensation (Ruzicka et al. (; Jacquet and Marrocchi 2017). The oscillatory sector zoning shown by the exceptional NWA 11086 IO12 (Fig. 5) also suggests departure from equilibrium, but the analogy with fast diffusion-controlled grown planetary igneous olivine (Welsch et al. (2014 is consistent with an igneous character. Adding to this the textural continuum between IOG and *bona fide* chondrules, we contend, in agreement with the latest literature (e.g. Jones et al. 2000; Pack et al. 2005; Russell et al. 2010; Ushikubo et al. 2012), that IOG are genetically related to chondrules.

4.2 Isolated olivine grains formed hot

It does not however follow that IOG result from the *cold* fragmentation of chondrules. Olsen and Grossman (1978) had already voiced skepticism about the feasibility to cleanly separate olivine phenocrysts from the mesostasis. While Richardson and McSween (1978) had invoked the friability of the altered mesostasis at least for CM chondrite chondrules, that alteration is now widely believed to have occurred on the parent body (Brearley (2014, too late for chondrule-chondrule collisions, and would not be relevant for the least altered chondrites such as ALH 77307. The large size of IOG compared to chondrule phenocrysts is also problematic if the former are fragments of the latter (e.g., Olsen and Grossman 1978). The larger IOG/chondrule ratio for type II compared to type I objects in carbonaceous chondrites would not be understandable if the chondrules were fragmented after being mixed together in the general protoplanetary disk.

The euhedral or equidimensional morphology of many IOG and their frequent concentric CL zoning patterns indicate that those feature unbroken olivine. Although, again, this study focuses on type I IOG, we recall that the ferroan IOG N6-7 in NWA 5958 exhibited oscillatory zoning visible in BSE (Jacquet et al. 2016; Jacquet and Marrocchi 2017). The frequent presence of near-complete enstatite margins (e.g. Fig. 3G,H; Fig. 4C,D) in type I IOG is comparable to those shown by whole chondrules. These are ascribed to the influx of SiO into residual melts, promoting pyroxene crystallization at the expense of olivine (e.g., Libourel et al. (; Chaussidon et al. (; Friend et al. (; Barosch et al. 2019). This indicates that the IOG experienced such influx as independent objects. In fact, prior to enstatite precipitation, the decrease of refractory incompatible elements toward IOG olivine margins reflected by the CL zoning also suggests condensation of Mg and SiO into the parent melt, inducing dilution of these elements, as advocated by Jacquet and Marrocchi (2017), as well as Marrocchi et al. (2018, 2019) and Libourel and Portail (2018) for the palisadic olivine. More volatile Fe, Mn, Cr would have recondensed contributing to their outward increase in the olivine. Perhaps in some cases—for those IOG which really were bare euhedral olivine grains—the late olivine condensed directly, without any intermediate liquid, which may account for the patterned surfaces which Olsen and Grossman (1974) compared to terrestrial sublimate olivine. In such cases, glass inclusions would of course have to date back to the time the olivine was still hosted in the partially molten chondrule. Thus IOG experienced gas-melt (or gas-solid) interaction as isolated objects. In that sense, condensation played a role, after all, in IOG formation.

Further evidence for a hot formation of IOG is afforded by compound objects, either between IOG (e.g. Fig. 3E,F) or with chondrules (e.g. Fig. 1G), which require collisions in a plastic state (Gooding and Keil (; Akaki and Nakamura 2005). Some *bona fide* chondrules with coarse olivine phenocrysts may have engulfed them during cooling—that is, be "enveloping" compound chondrules in the parlance of Wasson et al. (1995). To be sure, many IOG *are* mere misshapen olivine (± mesostasis, pyroxene) debris, but even those may sometimes be fragments of hot-formed IOGs (e.g., the IOG "half" NWA 5958-4 IO13; Fig. 3C,D), or have been fragmented hot, if enstatite has surrounded them since (e.g. NWA 5958-4 IO1 in the Electronic Annex). We note that Jones (1992) did envision isolated olivine grain formation from fragmentation of "macroporphyritic chondrules", although the formation of the latter was left unexplained.

We conclude from enstatite rimming and concentric CL zoning in many IOG (Fig. 3-4) that they formed as stand-alone high-temperature objects. In that sense, we deem it warranted to also call them "chondrules"(*sensu lato*, e.g. Jacquet et al. 2012; Jacquet and Marrocchi 2017), whenever the distinction is not of interest, as they are part of the same continuum of objects (with the word "chondrule" being already applied to a diverse suite of objects: ferromagnesian, aluminium-rich, chromite-rich, etc. see e.g. Connolly and Desch 2004). This does not exclude the possibility that some isolated olivine grains in chondrites have different origins, e.g. as AOA or a forsterite-bearing CAI fragment, although no evidence (e.g. isotopic) to that effect has been found for those of this study. To avoid any confusion, we remind the reader that the expression "*bona fide* chondrule" in this paper continues to refer to a non-IOG chondrule.

4.3 Isolated olivine grains as chondrule splashes

Since IOG are (by definition) olivine-dominated, they cannot be of chondritic composition, if only because of the Mg/Si (atomic) ratio (2 in endmember forsterite vs. 1.02 for solar abundances, Lodders (2003. This undercuts the Jacquet and Marrocchi (2017) effort to explain the IOG olivine incompatible element enrichment relative to "mainstream" type I chondrules in terms of varying olivine subtraction from a given parental melt. Most likely, the IOG precursors oversampled olivine. We must thus attempt to identify their nature.

Given the general analogy of chondrule relict grains with AOA olivine (Marrocchi et al. 2018, 2019), a first candidate would be simply AOAs. However, melting experiments of AOA analogs by Whattam et al. (2010) produced porphyritic textures. Also bulk AOA composition, although deficient in silica relative to type I chondrules, are closer to the latter than IOG (because of their CAI-like inclusions). Perhaps, though, a fragment of the olivine portion of a compact AOA (the dominant AOA texture in CM and related chondrites; Jacquet et al. (; Krot et al. (2004 would do. However, the (polycrystalline) olivine parts rarely extend beyond a few tens of μm in extent (e.g., Krot et al. (2004), at variance with the large IOG sizes. Furthermore, the abundance of AOAs overall in the studied chondrites is too low (~1 vol%) to account for the abundances of magnesian IOG (2-3 vol%). Indeed, the substantial matrix fraction in carbonaceous chondrites indicates that only a minority of their matter, and in particular of their AOAs, has been converted in chondrules (Marrocchi et al. 2019).

A second precursor candidate that comes to mind would be olivine phenocrysts from previous generations of chondrules. Now, if the chondrule/matrix ratio of carbonaceous chondrites, in particular CM-related chondrites, can be again taken as a measure of the extent of chondrule formation, chondrule-to-chondrule recycling must have been limited. Assuming chondrule-forming events were independent, Marrocchi et al. (2019) calculated that a fraction of 12 % of chondrule material was inherited from earlier chondrules (*sensu lato*) in NWA 5958 (the same formula would yield 19 % for Allende and 26 % for ALH 77307). At face value, this would, when multiplied by the chondrule abundance, approximate the modal abundance of IOG. However, it is unlikely that the recycled chondrules were preferentially in the form of free-floating olivine phenocrysts when they were remelted. This is because, again, fragmentation would not necessarily liberate bare olivine phenocryst (Olsen and Grossman 1978), and presumably, before the purported second chondrule-forming event, such olivine grains would have mixed with chondritic dust, such as rim high-temperature objects in CM chondrites (Metzler et al. (1992, erasing their olivine-dominated character most of the time.

Separating olivine crystals from the mesostasis would be conceivable if the mesostasis were *liquid*, that is during chondrule formation[2]. The crystals and liquid could separate (to some extent) owing to buoyancy during sudden accelerations. Such could be caused by disruptions under strong headwind (Kato et al. (2006 or more generically by chondrule-chondrule collisions, producing splashes of melt and crystals, with varying proportions thereof, the IOG representing the olivine-dominated end of the spectrum. Some crystals may have been fragmented, either during initial splashing or subsequent collisions, and the latest ones not rounded up by further olivine crystallization may account for the misshapen IOG (devoid of euhedral/subhedral shape and concentric CL). Collisions among hot chondrules must have been relatively frequent as a few percent of chondrules are compound (Gooding and Keil (; Wasson et al. 1995; Akaki and Nakamura 2005; Ciesla et al. (2004. A collisional origin during chondrule formation may explain the higher proportion of IOG among type II chondrules as the higher concentrations of solids inferred for their formation regions (e.g., Schrader et al. (; Tenner et al. (2017 Tenner et al. (2015 would promote collisions, although their rate also depends on relative velocities and the time in plastic state (which may have been shorter; Jacquet et al. (2015.

---

[2] In which case speaking of the olivine phenocrysts as "precursors" might be misleading as they would have formed during the same episode as the IOG.

Incidentally, the over-representation of type II chondrules among IOG might explain the dominance of ferroan olivine in Wild 2 terminal particles (13 vs. 8 (non-AOA-like) forsterites in the compilation of Defouilloy, Nakashima, Joswiak, Brownlee, at variance with type I chondrule dominance in carbonaceous chondrites. Indeed, if chondrule-forming regions did not extend to the accretion reservoir of this comet, the latter may have received some of their products by aerodynamic transport, but these would be biased toward the smaller fragments, less easily decoupled from the gas (Jacquet (2014.

Although the chemical and isotopic properties of IOG overlap with *bona fide* chondrules, their large size compared to the phenocrysts of the latter might question whether they derive from the same chondrules rather than some largely lost population. However IOG represent 5-13 % of type I chondrules *sensu lato* (Table 1), and if correctly interpreted as chondrule splashes, are only the tip of the iceberg: one would need to add isolated pyroxene grains, some cryptocrystalline chondrules (melt-dominated splashes) and larger fragments with more representative silicate/melt ratios, which, if subsequently relaxed to some extent to spherical shapes, would be indistinguishable from unsplashed chondrules. So the IOG source chondrules have to represent a few tenths at least of the whole type I chondrule inventory in carbonaceous chondrites. If we are to allow a significant population of non-IOG-related chondrules, we need to assume near-complete destruction of the IOG source chondrules, and very little splashing of the others, that is, very different collisional histories despite the thermal history similarity suggested by petrography. It seems simpler to assume that they formed in the same regions and that collisions induced about a tenth of the chondrule matter to be splashed out in the form of IOG.

If so, how can we then explain the large size of IOG? The PO chondrules studied by Marrocchi et al. (2018, 2019) and Libourel and Portail (2018) may offer a clue. Indeed, they are surrounded by thick (~100 μm) palisades of olivine that asymmetrically grew toward the exterior, presumably as a result of Mg and Si addition from the gas (Fig. 1C,D; Marrocchi et al. 2018, 2019). Now, an isolated olivine expelled from its parent chondrule would have interacted with the gas on all sides (as inferred in section 4.2), with no competing crystal around, so may well have hereby attained a diameter double of that of the palisadic olivine grains, about as observed (Fig. 2). Roughly speaking, the refractory (bright CL) core of the IOG might correspond to the olivine crystallized in the parental chondrule and the margin (if concentric) to that formed after splashing, but diffusion (for Ca) after splashing as well as recondensation onto the chondrule before would make the actual boundary between these two stages uncertain.

Later, as mentioned in the previous subsection, the olivine would have reacted with SiO and have been replaced by enstatite to varying extent. Some objects may hence have lost their olivine-dominated nature: this may explain the largest isolated pyroxene grains seen by Jones (1999) in ALH 77307 even though some isolated pyroxene grains may be merely surficial sections of pyroxene-armoured IOG. Yet others could of course have been expelled as pyroxene grains, after pyroxene crystallization had commenced in chondrules. At any rate, Jones (1999) deemed derivation of isolated pyroxene grains from chondrules likely from textural and mineral chemical similarities with them but a dedicated study beyond her work only published in abstract form would be desirable.

The emerging scenario of IOG formation is sketched in Fig. 12.

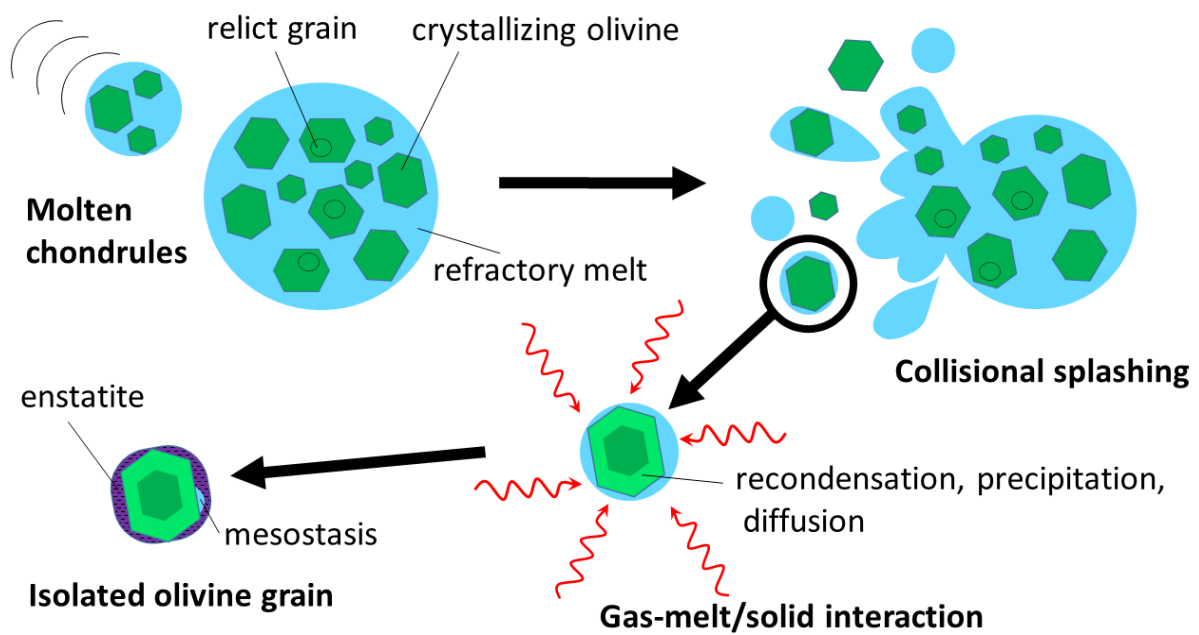

**Figure 12**: Sketch of the proposed formation scenario for IOG. Chondrule collisions produce splashes, some dominated with one olivine phenocryst, which undergo interaction with the gas upon cooling.

4.4 Refractory forsterites and diversity among type I chondrules in carbonaceous chondrites

We have yet to understand refractory forsterites that is, those CaO-rich IOG studied by e.g. Steele (1988); Weinbruch et al. (2000), Pack et al. (2005). CaO enrichment is also seen in the above mentioned palisadic olivine grains in porphyritic olivine chondrules, which certainly would not have been the least prone to expulsion upon collisions. Marrocchi et al. (2018, 2019) explained the Ca, Al, Ti-enrichment in those by precipitation out of a Ca-Al-Ti-rich melt. Indeed, in a reservoir with dust/gas ratio below unity, the melt upon olivine crystallization should be fairly refractory because of evaporation (Ebel and Grossman (; Ruzicka et al. (2008. The palisadic olivine grains would have escaped diffusional resetting of their Ca after recondensation because of their size, the same would hold *a fortiori* for the large isolated refractory forsterites. This incidentally constrains the diffusion length of Ca to a few tens of microns, hence the timescale of days inferred by Marrocchi et al. (2018, 2019).

It remains to be understood why high refractory element contents are restricted to the $^{16}$O-rich ($\Delta^{17}$O < -4 ‰) population of IOG or chondrules at large, as observed by Jacquet and Marrocchi (2017). These authors inferred protracted cooling to allow equilibration of the olivine with a late, incompatible-element-enriched melt, but their closed-system olivine crystallization reasoning failed to recognize, as above, that the *initial* melt was incompatible-element-rich. Ushikubo et al. (2012), Schrader et al. (2013) and Tenner et al. (2015, 2017) noted that $^{16}$O-poor chondrules tended to have more ferroan olivine than the $^{16}$O-rich counterparts, which they ascribed to greater solid/gas ratios in the chondrule-forming regions (with the solids (dust and ice) being $^{6}$O-depleted). $^{16}$O-poor type I IOG do not, however, present such a FeO enrichment, but their large size may have prevented diffusion of Fe$^{2+}$ into their core during cooling. A higher solid/gas ratio would lead to lower refractory element concentrations in olivine at the onset of its crystallization and an upturn in the CaO vs. FeO trend (e.g. Ebel and Grossman 2000), as observed (Fig. 8A).

In type I porphyritic chondrules, moderately high fayalite contents (above ~2 mol%) seem correlated with pyroxene proportion: the three PP chondrules in CR3 chondrites studied by Tenner et al. (2015) have Mg# $\equiv$ 100 × Mg/(Mg+Fe) of 94.2-97.8, with no PO in this range although it comprises a majority of their type I chondrules; the four PP chondrules of Lewis Cliff 85332 (C3-ung, CR-related) analysed by Wasson et al. (2000) have Fs$_{4.2-9}$. Outside carbonaceous chondrites, type IAB and IB chondrules in Semarkona (LL3.0) span Fa$_{1.8-6.7}$ in the study of Jones (1994). On the other hand, no Mg# below 98.4 (nor $\Delta^{17}$O > -3.55 ‰) is reported in any type I chondrule from the CV3 Kaba and NWA 8613 analysed by Hertwig et al. (2018 Hertwig , although they include POP and PP chondrules; same for the Murchison (CM2) chondrules studied by Chaumard et al. (2018), save for one granular olivine pyroxene (GOP) chondrule with Mg# = 96 ($\Delta^{17}$O = -2.7 ± 0.3 ‰). The situation of the Yamato 81020 (CO3.0) study by Tenner et al. (2013) is more marginal, with the 3 PP having Mg# of 97.8-99 (all $^{16}$O-rich), while all 3 type I chondrules with Mg# < 97 are POP and $^{16}$O-poor. Of course, the estimate of the true proportions of pyroxene is hampered by sectioning artifacts (Hezel and Kießwetter (2010. Still, if there is indeed some link between "ferroan forsterite" and pyroxene abundance, this may indicate that the chondrules in question cooled more slowly, since pyroxene should appear at lower temperature than olivine (e.g. Ebel and Grossman 2000). Indeed, in Vigarano, the most pyroxene-rich chondrules seem to have the lowest proportion of clinoenstatite in low-Ca pyroxene, indicative of slower cooling at least around 1000 °C (Brearley et al. (; Soulié (2014.

Thus, the $^{16}$O-rich type I chondrules and the associated refractory forsterites may have cooled more rapidly, under lower solid/gas ratios, than their generally more ferroan $^{16}$O-poor counterparts. Are these two subtypes of type I chondrules derived from different sources? Or were two styles of type I chondrule formation events overlapping in individual reservoirs? The $^{16}$O-poor variety is more prevalent in CR chondrites than in other carbonaceous chondrites: Keeping $\Delta^{17}O = -4$ ‰ as our arbitrary boundary between the two, Tenner et al. (2015) found a 10:35 ratio between $^{16}$O-rich and $^{16}$O-poor type I chondrules in CR3 chondrites; the same research group obtained a ratio of 19:4 in the CO3.0 chondrite Y 81020 (Tenner et al. 2013), 23:2 in the CM2 chondrite Murchison (Chaumard et al. 2018) and 71:7 in CV chondrites (Rudraswami et al. (; Hertwig et al. (; Hertwig et al. (2019. One could envision that $^{16}$O-poor type I chondrule formation was merely more frequent in the CR chondrite reservoir without denying $^{16}$O-poor type I chondrule formation elsewhere. However, $^{16}$O-poor chondrules appear systematically more $^{54}$Cr-rich than their $^{16}$O-rich counterparts in both CR and CV chondrites (e.g., Williams et al. 2020; Schneider et al. (under review)). This indicates a different reservoir of origin for the $^{16}$O-poorer type I chondrule, which presumably was spatially and/or temporally closer to the accretion event of CR chondrites. We note that CR chondrules exhibit initial $^{26}$Al abundances lower than their counterparts in CO and CV chondrules (Nagashima et al. 2014; Schrader et al. 2017; Tenner et al. 2019), but possibly comparable to cometary samples (Ogliore et al. 2012; Nakashima et al. 2015). The Fe/Mn ratio of olivine in type II chondrules, lower in CR chondrites than in other carbonaceous chondrites (e.g., Berlin et al. 2011; Jones (; Frank et al. 2014) also sets the CR chondrule population apart among carbonaceous chondrites.

So the emerging picture is that of a CR-like chondrule population, with apparently higher solid/gas ratios and slower cooling than the others (for the type I chondrules), which contributed to non-CR reservoirs dominated by $^{16}$O-rich chondrules. Such a limited mixing of chondrules beyond their formation region could be reproduced in the simulations of Goldberg et al. (2015).

This limited mixing would also have involved refractory inclusions, as suggested by Al-Mg isotopic evidence (Larsen et al. (2020). While the $^{16}$O-poor type I chondrules of different carbonaceous chondrite groups (CO, CV, CM) may thus come from a single source, the $^{16}$O-rich ones, although O isotopically similar, may still have diverse origins, as suggested by distinctive petrographic features in the different chemical groups (e.g. size, prevalence of fine- or coarse-grained rims, occurrence of primary feldspar… see Jones 2012).

## 5. Conclusion

We have carried out microscopic, cathodoluminescence, chemical and O isotopic measurements of type I isolated olivine grains (IOG) in the carbonaceous chondrites Allende (CV3), Northwest Africa 5958 (C2-ung), Northwest Africa 11086 (CM-an), Allan Hills 77307 (CO3.0).

The IOG typically represent a few percent of the studied carbonaceous chondrites, and are dominated by type I IOG but with an overrepresentation of type II compositions. The type I IOG present O isotopic signatures similar to *bona fide* chondrules, although they are generally coarser than chondrule phenocrysts. CaO, which may reach up to 0.9 wt% in $^{16}$O-rich IOG, as well as $Al_2O_3$ and $TiO_2$ decrease from core to rim, while FeO, MnO, $Cr_2O_3$ increase. Electron microprobe traverses and CL imaging often reveal concentric zoning, with enstatite frequently rimming the olivine's outer edge.

The general isotopic and chemical similarities, indeed the textural continuum, with *bona fide* chondrules indicate that the IOG were derived from them. Their not uncommonly unbroken morphology and evidence of interaction with the gas, with recondensation on all sides possibly accounting for their large size, indicates that rather than being cold fragments of chondrules, they were expelled from chondrules while these were molten, likely during chondrule-chondrule collisions. Among IOG, the refractory forsterites retained their high-temperature composition presumably established by equilibration with a refractory melt, as some palisadic olivine grains in porphyritic olivine chondrules. These $^{16}$O-rich type I chondrules may have undergone quicker quenching than their $^{16}$O-poorer (type I) counterparts, possibly derived from a CR chondrite-like reservoir.

Thus, IOG were likely derived from chondrules, as favored by most recent authors (e.g. Jones 1992; Jones et al. 2000; Pack et al. 2005; Russell et al. 2010), but were affected by significant gas-solid interactions (before and after expulsion from the parent chondrules), as inferred by their earliest students (e.g. Fuchs et al. 1973; Olsen and Grossman 1974, 1978; Steele 1986, 1988, 1989), hereby reconciling the two endmembers of the isolated olivine literature.

*Acknowledgments*: We thank the associate editor and two anonymous reviewers for their extensive reviews which substantially improved the readability of the discussion. We are grateful to the Muséum national d'Histoire naturelle de Paris for the NWA 5958 and Allende samples. The MNHN gives access to the collections in the framework of the RECOLNAT national Research Infrastructure. We thank the "Meteorite Working Group" at the Johnson Space Center for the loan of the ALHA 77307 sample. US Antarctic meteorite samples are recovered by the Antarctic Search for Meteorites (ANSMET) program which has been funded by NSF and NASA, and characterized and curated by the Department of Mineral Sciences of the Smithsonian Institution and Astromaterials Curation Office at NASA Johnson Space Center. This research was funded by the Agence Nationale de la Recherche (ANR) through Grant ANR-14-CE33-0002-01 SAPINS (Principal Investigator Yves Marrocchi.), ANR-18-CE31-0010-01 CASSYSS (co-Investigator Yves Marrocchi.) and ANR-15-CE31-0004-1 CRADLE (co-Investigator Emmanuel Jacquet). This article is dedicated to the memory of Nicole Guilhaumou.